\documentclass[twocolumn,aps,prl,showpacs,preprintnumbers,floatfix]{revtex4}
\usepackage{amssymb}

\usepackage{graphicx}
\usepackage{dcolumn}
\usepackage{bm}
\usepackage{epsfig}
\usepackage{amsmath}

\begin{document}

\title{NMR Evidence for Gapped Spin Excitations in Metallic Carbon Nanotubes}
\author{P.M. Singer,$^{1}$ P. Wzietek,$^{1}$ H. Alloul,$^{1}$ F. Simon,$^{2}$, and
H. Kuzmany $^{2}$}
\affiliation{$^{1}$ Laboratoire de Physics des Solides, UMR 8502,
Universit\'{e} Paris-Sud, 91405 Orsay, France}
\affiliation{$^{2}$ Institut
f\"{u}r Materialphysik, Universit\"{a}t Wien, Strudlhofgasse 4, A-1090 Wien,
Austria }
\date{\today}

\begin{abstract}
We report on the spin dynamics of $^{13}$C isotope enriched inner-walls in double-wall carbon nanotubes (DWCNT) using $^{13}$C nuclear magnetic resonance (NMR). 
Contrary to expectations, we find that our data set implies that the spin-lattice relaxation time ($T_{1}$) has the same temperature ($T$) and magnetic field ($H$) dependence for most of the innerwall nanotubes detected by NMR.
In the high temperature regime ($T \gtrsim 150$ K), we find that the $T$ and $H$ dependence of $1/T_{1}T$ is consistent with a 1D metallic chain. For $T \lesssim 150$ K we find a significant increase in $1/T_{1}T$ with decreasing $T$, followed by a sharp drop below $\simeq$ 20 K. The data clearly indicates the formation of a gap in the spin excitation spectrum, where the gap value $2\Delta \simeq $ 40 K ($\equiv$ 3.7 meV) is $H$ independent.
\end{abstract}
\pacs{71.20.Tx, 61.46.+w, 73.22.-f, 76.60.-k}

\maketitle

\preprint{APS/revtex4}

The electronic properties of carbon nanotubes have been of a topic of intense
investigation ever since their discovery in early 1990's. According to
band-structure calculations the basic electronic structure of single-wall
carbon nanotubes (SWCNT) is expected to depend on the chiral wrapping
vector $(n,m)$ across the graphene plane, where tubes for which $(2n+m)/3=$ 
\textit{integer} are metallic, while all other tubes are semiconducting \cite
{mintmire,saito,dress,hamada} with a large $\sim $1 eV gap \cite{white}. While STM and
transport measurements on \textit{isolated} tubes demonstrate the diversity
of tube properties, significant measurements on \textit{macroscopic} amounts
of tubes are only possible in selected cases. Photoemission spectroscopy
(PES) measurements on metallic tubes in bundles \cite{ishii,rauf} suggest
that strong electron-electron correlations can lead to a
Tomonaga-Luttinger-liquid (TLL) state. Recently, double wall carbon nanotubes (DWCNT)
have been synthesized by filling SWCNT with fullerenes (so called ''peapods''\cite{smith}) followed by a high temperature reaction which merges the fullerenes into inner tubes \cite{luzzi,bandow}. These DWCNT have some exceptional properties since the inner tubes are accommodated in a highly shielded environment under clean room conditions \cite{pfeiffer}. Raman experiments performed
even on bucky paper material allows one to detect some significant properties of these
inner tubes due to their small diameter (high curvature).

Nuclear magnetic resonance (NMR) is usually an excellent technique for probing
the electronic properties at the Fermi level of metallic systems, take for instance conducting polymers, fullerenes, and high temperature superconductors to mention a few. However
the 1.1\% natural abundance of $^{13}$C with nuclear spin $I$=1/2
limits the sensitivity of such experiments. Data taken on SWCNT essentially
evidence a large distribution of properties since samples of identical tubes
are presently out of reach. 
In this report, selective enrichment of the inner shells using $^{13}$C isotope enriched fullerenes \cite{simon,simon2} in the 
"peapod" synthesis route is used to probe the specific properties of the inner tubes. The $^{13}$C enrichment allows us to
increase the $^{13}$C NMR sensitivity by two orders of magnitude, and furthermore
achieve \textit{selective} $^{13}$C enrichment of the inner shells alone. 
This provides us with the possibility of singling out the
electronic properties of these inner tubes for the first time. 
We show that, although these tubes are distributed in
diameter and chirality, their electronic properties display a strikingly
homogeneous behaviour. The magnetic properties of these inner wall
nanotubes behave as for a 1D metal at room $T$, but exhibits a pronounced gap
below $\simeq $20 K. This unexpected result reveals
that this specific macroscopic collection of carbon nanotubes is an object
displaying original physical properties worth studying in more detail
with macroscopic experimental techniques.  

All $^{13}$C NMR data in this report were taken with the sample sealed in a 6 mm diameter
glass tube filled with 200 mbar of high purity Helium gas. Details of the
synthesis techniques and various independent experimental evidences for the formation of 89\% $^{13}$C isotope enriched inner-walls inside natural 1.1\% $^{13}$C enriched outer walls are reported elsewhere \cite{simon}. We probed the low frequency spin dynamics (or low energy spin excitations, equivalently) of the inner-tubes using the spin
lattice relaxation time, $T_{1}$, defined as the characteristic time it takes the $^{13}$C nuclear
magnetization to recover after saturation. The signal intensity after saturation, $S(t)$, was
deduced by integrating the fast Fourier transform of half the
spin-echo for different delay times $t$. All data were taken
with excitation pulse lengths $\pi/2 = 3.0 \mu$s and short pulse separation
times of $\tau=15 \mu$s  \cite{slichter}. We obtained the value of $T_{1}$ by fitting the $
t$ dependence of $S(t)$ to the form $
S(t) \ = \ S_{a} - S_{b} \cdot M(t)$, where $S_{a}  \simeq S_{b}$ ($>0$) are arbitrary signal amplitudes, and 
\begin{equation}
M(t) = \exp\left[-\left(t/T_{1}^{e}\right)^{\beta}\right]
\label{stretch}
\end{equation}
is the reduced magnetization recovery of the $^{13}$C nuclear spins.
Fig. \ref{recov} shows the results of $M(t)$ for the inner-tubes as a function of the scaled delay
time $t/T_{1}^{e}$, under various experimental conditions listed in the Figure. We find that $M(t)$ does not follow the single
exponential form with $\beta=1$ (dashed line), but instead fits well to the
stretched exponential form with $\beta \simeq 0.65(5)$ which implies a distribution in underlying relaxation times $T_{1}$ across the sample. In such cases, $T_{1}^{e}$ in Eq. (\ref{stretch}) is directly proportional to the mean value $\overline{T_{1}}$ of the 
$T_{1}$ distribution as such $T_{1}^{e} = \overline{T_{1}} \cdot \beta / \Gamma(1/\beta)$, where $\Gamma$ is the gamma function.
We display the data in Fig. \ref{recov} on a semi-log scale for the $time$ axis in order to
accentuate the data for earlier decay times and to illustrate the collapse
of the data set for the upper 90\% of the NMR signal. We find that the upper 90\% of the $M(t)$ data is consistent with constant $\beta \simeq 0.65(5)$ (see inset), implying a constant underlying distribution in $T_{1}$ for a large range of experimental conditions. The lower 10\% of the $M(t)$ data (corresponding to longer delay times) comes from the non-enriched outer-walls 
which, as a result of their larger diameters, have much longer relaxation times under similar experimental
conditions \cite{tang,goze,shimoda,kleinhammes}.

Two distinct origins for the multi-exponential magnetization recovery can be
considered. The first is due to the powder average over the spatial
anisotropy in $T_{1}$. The distribution is independent of the tube
properties, and can also be found in the $^{13}$C NMR data for alkali doped
fullerenes A$_{n}$C$_{60}$ \cite{tycko,brouet}. Given the similar diameter
of C$_{60}$ ($d=$ 0.71 nm) to the average inner-wall diameter ($\overline{d}=$ 0.7
nm \cite{simon,simon2}) in this report, we can expect comparable bonding effects for the
electron orbitals. It has been shown that in A$_{n}$C$_{60}$ the $
T_{1}$ for $^{13}$C is dominated by dipole-dipole interactions between the
electron spin in the $pp\pi $ bond and the $^{13}$C nuclear spin \cite
{antropov}. In this case, the relaxation depends on the orientation of the $
p_{\pi}$ orbital (which is perpendicular to the tube surface) and the
external magnetic field, and therefore contributes to the multi-exponential
form of magnetization recovery for a powder average. This resultant $T_{1}$ distribution
is independent of $T$ and $H$.

Another source of multi-exponential recovery is from a distribution of the
inner tube properties themselves, such as their diameter. According to Raman
scattering, the inner tubes have a mean diameter of $\overline{d}\simeq 0.7$
nm with a standard deviation of $\sigma \simeq $ 0.1 nm \cite{simon,simon2}.
Within this distribution lies a variety of tubes with different chirality
and one can {\it a priori} expect to find metallic as well as
semiconducting tubes \cite{dress}. If both semiconducting and metallic inner-tubes existed in our sample, one would expect the ratio of the $T_{1}$'s between the different tubes to increase exponentially with decreasing $T$ below the semiconducting gap ($\sim 5000$ K \cite{white}), which would drastically change the underlying $T_{1}$ distribution with decreasing $T$. This change would manifest
itself as a large change in the shape of the recovery $M(t)$, however, as shown in Fig. \ref{recov} this is $not$ the case. We can therefore rule out the possibility of two components in $T_{1}$ with
different $T$ dependences, and instead we conclude {\it that all $T_{1}$ components exhibit the same $T$ and $H$ dependence within experimental scattering.}

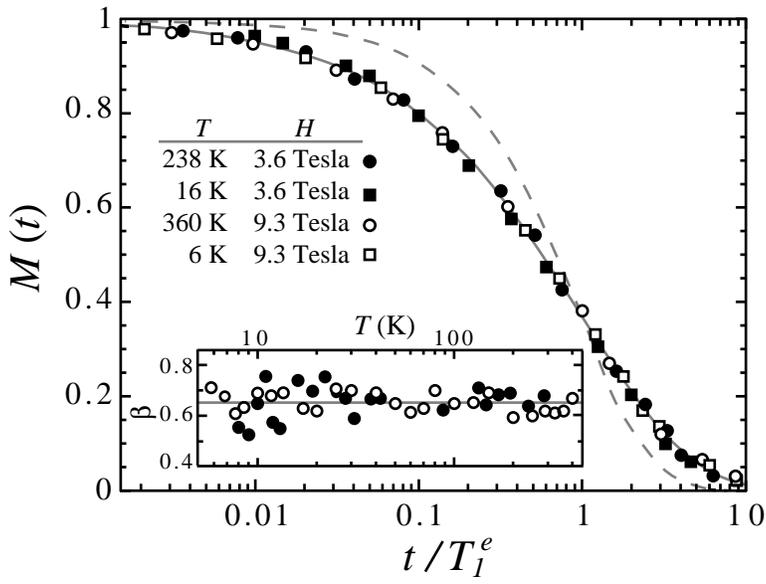
\begin{figure}
\caption{Temperature dependence of spin-lattice relaxation rate divided by temperature, $ 1/T_{1}^{e}T$, in units of $(10^{3}\times$ s$^{-1}$ K
$^{-1}$). Grey curves are best fits to Eq. (\ref{gapT1}) with $2\Delta = 46.8 (40.2)$ K for $H= 3.6 (9.3)$ Tesla, respectively.}
\label{T1T}
\end{figure}

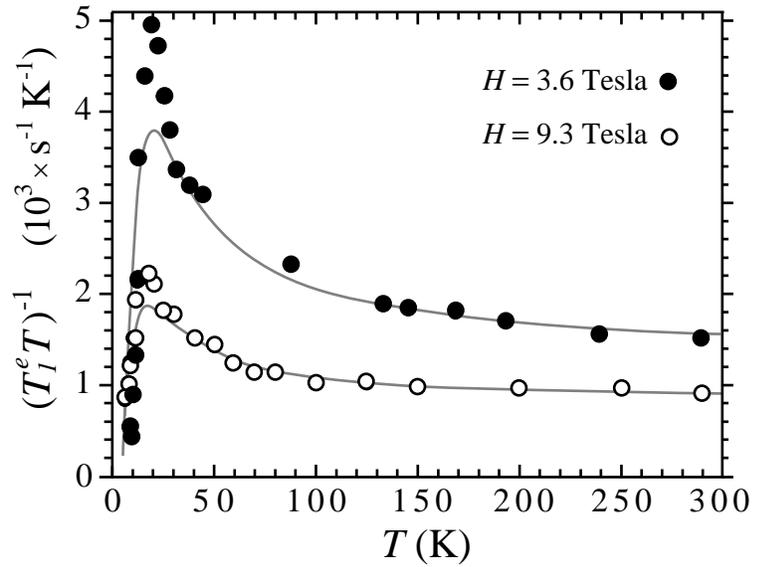
\begin{figure}
\caption{$ 1/T_{1}^{e}T$, in units of $(10^{3}\times$ s$^{-1}$ K
$^{-1}$), at fixed $T= 290$ K, plotted as a function of $1/\protect\sqrt{H}$, in units of 
(Tesla$^{-1/2}$). Linear fit corresponds to 1/$T_{1}^{e}T = A+B / \protect
\sqrt{H}$ with B = 0.00206 (Tesla$^{1/2}$ s$^{-1}$ K$^{-1}$) and A =
0.00028 (s$^{-1}$ K$^{-1}$).}
\label{sqrtH}
\end{figure}

The $T_{1}$ distribution in the sample, whether it arises from anisotropy or diameter variations (or both), shows a uniform $T$ and $H$ dependence. It is therefore appropriate to follow the $T$ and $H$ dependence of the mean value of the distribution ($T_{1}^{e}$ in Eq. (\ref{stretch})), and thereby get insight into the homogenous electronic state of the inner tubes.
In order to avoid unnecessary experimental scattering in $T_{1}^{e}$, we then go back and fit all the $M(t)$ data to
Eq. (\ref{stretch}) with a $fixed$ value of $\beta = 0.65$.
We plot the resulting temperature dependence of $1/T_{1}^{e}T$ in Fig. \ref{T1T} for two different values of the magnetic field $H$. We can immediately separate the data into two temperature regimes; the high
temperature regime $\gtrsim$ 150 K, and the low $T$ regime $\lesssim$ 150 K. At high
temperatures we find that $1/T_{1}^{e}T$ is independent of $T$ which indicates a metallic state \cite{slichter}, which given the arguments above implies that all of the inner tubes are metallic.
We also find a strong field dependence for $
T_{1}$ which is best illustrated by plotting the high temperature value of $1/T_{1}^{e}T$ against $1/
\sqrt{H}$ for $H$ values ranging from 1.2 Tesla to 9.3 Tesla, as shown in
Fig. \ref{sqrtH}. We find that the data fit well to the form 
\begin{equation}
\frac{1}{T_{1}^{e}T} \ = \ A + B \frac{1}{\sqrt{H}} ,  \label{sqrt}
\end{equation}
where $A$ and $B$ are constants, which is very suggestive of a 1D spin
diffusion mechanism for $T_{1}$ \cite{boucher,soda,bourb}. $B/\sqrt{H}$ corresponds to
the diffusive contribution to the relaxation originating from the long
wavelength (i.e. $q \simeq 0$) modes, while $A$ corresponds to the
non-diffusive contributions from $q > 0$ modes. A cutoff to the divergence
in Eq. (\ref{sqrt}) as $H \rightarrow 0$ is often encountered in 1D spin
chain systems \cite{boucher,soda,bourb} due to inter-chain coupling. In the present
case, we can postulate that electron tunneling between inner to outer walls
could cause similar cutoff effects, however the data down to the lowest
field of $H$=1.2 Tesla indicates that the cutoff has not been reached yet.
We therefore conclude that the high-temperature regime is consistent with a 1D metallic chain.

The origin of the unusual $T$
dependence of 1/$T_{1}^{e}T$ in the low temperature regime ($\lesssim 150$ K) is not immediately obvious. We can however rule
out certain possibilities. Firstly, we can rule out the possibility of an
activation type mechanism where $T_{1}$ is dominated by fluctuating
hyperfine fields which are slowing down with decreasing $T$. If this were the case the 
temperature where 1/$T_{1}^{e}T$ reached its maximum would shift with the resonance frequency $\omega $ \cite{slichter}, or
the applied magnetic field $ \omega =\gamma _{n}H$ ($\gamma _{n}/2\pi
=10.71$ MHz/Tesla is the gyromagnetic ratio for $^{13}$C), equivalently. As shown in Fig. 
\ref{T1T}, however, we find no evidence of a shift in the peak temperature
with $H$. Furthermore, at low temperatures 1/$T_{1}^{e}T$ is found to drop below its
high temperature value which rules out the possibility of an activation contribution $plus$ a $T$ independent contribution.
Secondly, we also rule out the possibility that the $T$ dependence of 1/$T_{1}^{e}T$ is a result of paramagnetic centers which can arise from wall defects or impurity spins. The fact that a pronounced $gap$ exists in 1/$T_{1}^{e}T$ implies a pronounced gap in the low energy spin excitation spectrum, which cannot be explained by the presence of paramagnetic centers. We note that at the lowest temperatures $< 5$ K (not shown), $T_{1}^{e} $ becomes so long $(> 300$ s) that the low energy spin excitations specific to the homogeneous properties of the inner-tubes become inefficient, and other excitations take over, possibly defect related. In such cases we find that the shape of $M(t)$ is no longer universal and that the underlying distribution in $T_{1}$ is smoothed out, possibly as a result of nuclear spin-diffusion.

Having ruled out the above possibilities, we are then lead to consider the simplest 
explanation for the experimental data using a non-interacting electron model
of a 1D semiconductor with a small secondary gap (SG). The SG may be a
result of the finite inner-wall curvature \cite{hamada,kane,white,zol}, or
perhaps the applied magnetic field itself \cite{Lu}. We can fit the 1/$
T_{1}^{e}T$ data using this non-interacting model with only one free
parameter, the homogeneous SG, 2$\Delta $. We start by taking the normalized
form of the gapped 1D density-of-states $n(E)$
\begin{equation}
n(E)\ =\ 
\begin{cases}
\frac{E}{\sqrt{E^{2}-\Delta ^{2}}} & \text{for }|E|>\Delta \cr0 & \text{
otherwise}\cr
\end{cases}
\label{newDOS}
\end{equation}
also known as the van-Hove singularity ($E$ is taken with respect to the
Fermi energy). We then use Eq. (\ref{newDOS}) to calculate 1/$T_{1}^{e}T$ 
\cite{moriya} as such 
\begin{equation}
\frac{1}{T_{1}^{e}T}\ =\ \alpha (\omega )\int_{-\infty }^{\infty
}n(E)n(E+\omega )\left( -\frac{\delta f}{\delta E}\right) dE,  \label{gapT1}
\end{equation}
where $E$ and $\omega$ are in temperature units for clarity, $f$ is the Fermi function $f = [\exp(E/T) + 1 ]^{-1}$, 
and the amplitude factor $\alpha(\omega) = A + B'/\sqrt{\omega}$ is taken directly from Eq. (\ref{sqrt}) and 
Fig. \ref{sqrtH} (where $B' = 4.53\cdot10^{-5} $s$^{-1} $K$^{-1/2}$ for $\omega \equiv H$ in temperature units). We note that factoring out the diffusion effects from the integral in Eq. (\ref{gapT1}) is an approximation valid only if $A$
and $B$ are $T$ independent. Eq. (\ref{gapT1}) cannot be solved analytically, therefore we resort to numerical integration.
The results of the best fit to Eq. (\ref{gapT1}) are presented
in Fig. \ref{T1T}, where $2\Delta = 43(3)$ K ($\equiv$ 3.7 meV) is found to be $H$
independent (within experimental scattering) between 9.3 and 3.6 Tesla. We note that at the largest external field of 9.3 Tesla, $\omega =4.5$ mK $
\ll \Delta ,T$, however, $\omega $ must be retained inside the integral.
This is a consequence of the one dimensionality which yields a logarithmic divergence inside the integral of the form ln$(T/\omega)$ for $T \lesssim \Delta$.

{\it What could possibly be the origin of the observed gap}? Tight binding calculations
predict that applied magnetic fields can induce SG's of similar magnitude
for metallic SWCNT \cite{Lu}. However, such a scenario is excluded here from
the absence of field dependence of the observed gap. Our data would be more
consistent with a curvature induced SG for metallic tubes \cite
{hamada,kane,white,zol}, however for our typical inner-tubes the predicted
values, $\sim 100$ meV, are over an order of magnitude larger than our
experimental data. Other scenarios, such as quantization of levels due to
finite short lengths of the nanotubes could be considered as well, however, in
all these cases a behaviour independent of tube size and chirality
is certainly not expected.

This leads us to consider the effect of electron-electron interactions for
the metallic inner tubes. It has been predicted that electron-electron
correlations and a TLL state leads to an increase in 1/$T_{1}T$ with
decreasing $T$ \cite{yoshioka}, which is a direct consequence of the 1D
electronic state. The correlated 1D nature may also lead to a Peierls
instability \cite{dress} with the opening of a small collective gap $2 \Delta$ and a sharp drop in
1/$T_{1}T$ below $\Delta \sim 20$ K. Therefore, the
presence of both a TLL state \textit{and} a Peierls instability could
possibly account for the data, although here again, the independence on
tube geometry should be accounted for.

In conclusion, we have shown that the $T_{1}$ recovery data indicate that
most of the inner-tubes have similar $T$ and $H$ dependences, with no indication
of a metallic/semiconductor separation due to chirality distributions. At
high temperatures ($T\gtrsim 150$ K) 1/$T_{1}^{e}T$ of the inner tubes
exhibit a metallic 1D spin diffusion state, with no low-field cutoff down to
1.2 Tesla. This metallicity could result from charge transfer from the outer to the
inner tubes, however this speculation ought to be confirmed by independent experiments and theoretical calculations.
Below $\sim $150 K, 1/$T_{1}^{e}T$ increases dramatically with
decreasing $T$, and a gap in the spin excitation spectrum is found below $
\Delta \simeq $ 20 K. We list various interpretations for this temperature dependence, ranging from a non-interacting
secondary band-gap model to a 1D correlated electron model with a collective gap (possibly a Peierls instability). 
Firstly, these results should stimulate further experimental investigations on
diversely synthesized DWCNT in order to check whether these observations are specific
to the ''peapod'' synthesis route. Secondly, theoretical work on the incidence of 1D correlation effects for inner-wall nanotubes inside DWCNT should be helpful in sorting out the origin of our astonishing experimental evidence.

\section{Acknowledgments}

\begin{acknowledgments}
Support from the EU projects HPMF-CT-2002-02124, BIN2-2001-00580, and MEIF-CT-2003-501099, and the
Austrian Science Funds (FWF) project Nr. 17345, are recognized. The authors also wish to thank V. Z\'{o}lyomi and J. K\"{u}rti for valuable discussions.
\end{acknowledgments}

\clearpage

\bibliographystyle{apsrev}

\begin{thebibliography}{999}
\bibitem{mintmire}J. W. Mintmire {\it et al.}, Phys. Rev. Lett. {\bf 68}, 631 (1992).
\bibitem{saito}R. Saito {\it et al.}, Phys. Rev. B {\bf 46}, 1804 (1992).
\bibitem{dress}R. Saito, G. Dresselhaus, and M. Dresselhaus, {\it Physical Properties of Carbon Nanotubes} (Imperial College Press, 1998).
\bibitem{hamada}N. Hamada {\it et al.}, Phys. Rev. Lett. {\bf 68}, 1579 (1992).
\bibitem{white}J. W. Mintmire and C. T. White, Phys. Rev. Lett. {\bf 81}, 2506 (1998).
\bibitem{ishii}H. Ishii {\it et al.}, Nature {\bf 426}, 540 (2003).
\bibitem{rauf}H. Rauf {\it et al.}, Phys. Rev. Lett. {\bf 93}, 96805 (2004).
\bibitem{smith}B. W. Smith {\it et al.}, Nature (London) {\bf 396}, 323 (1998).
\bibitem{luzzi}B. W. Smith and D.E. Luzzi, Chem. Phys. Lett. {\bf 321}, 169 (1999).
\bibitem{bandow}S. Bandow {\it et al.}, Chem. Phys. Lett. {\bf 384}, 320 (2004).
\bibitem{pfeiffer}R. Pfeiffer {\it et al.}, Phys. Rev. Lett. {\bf 90}, 225501 (2003).
\bibitem{simon}F. Simon {\it et al.}, Phys. Rev. Lett. {\bf 95}, 17401 (2005).
\bibitem{simon2}F. Simon {\it et al.}, Phys. Rev. B {\bf 71}, 165439 (2005).
\bibitem{slichter}C. P. Slichter, {\it Principles of Magnetic Resonance} (Springer-Verlag, New York, 1989), 3rd ed.
\bibitem{tang}X. -P. Tang {\it et al.}, Science {\bf 288}, 492 (2000).
\bibitem{goze}C. Goze-Bac {\it et al.}, Carbon {\bf 40}, 1825 (2002).
\bibitem{shimoda}H. Shimoda {\it et al.}, Phys. Rev. Lett. {\bf 88}, 15502 (2002).
\bibitem{kleinhammes}A. Kleinhammes {\it et al.}, Phys. Rev. B {\bf 68}, 75418 (2003).
\bibitem{tycko}R. Tycko {\it et al.}, Phys. Rev. Lett. {\bf 68}, 1912 (1992).
\bibitem{brouet}V. Brouet {\it et al.}, Phys. Rev. B {\bf 66}, 155122 (2002).
\bibitem{antropov}V. Antropov {\it et al.}, Phys. Rev. B {\bf 47}, R12373 (1993).
\bibitem{boucher}J.-P. Boucher {\it et al.}, Phys. Rev. B {\bf 13}, 4098 (1976).
\bibitem{soda}G. Soda {\it et al.}, J. Phys. (Paris) {\bf 38}, 931 (1977).
\bibitem{bourb}C. Bourbonnais {\it et al.}, Phys. Rev. B {\bf 44}, 641 (1991).
\bibitem{kane}C. L. Kane and E. J. Mele, Phys. Rev. Lett. {\bf 78}, 1932 (1997).
\bibitem{zol}V. Z\'{o}lyomi and J. K\"{u}rti, Phys. Rev. B {\bf 70}, 85403 (2004).
\bibitem{Lu}J. P. Lu, Phys. Rev. Lett. {\bf 74}, 1123 (1995).
\bibitem{moriya}T. Moriya, J. Phys. Soc. Jpn.  {\bf 18}, 516 (1963).
\bibitem{yoshioka}H. Yoshioka, J. Phys. Chem. Solids {\bf 63}, 1281 (2002).
\end{thebibliography}

\begin{figure}
\caption{Reduced nuclear magnetization recovery, $M(t)$, as a function of the scaled delay
time $t/T_{1}^{e}$ (see Eq. (\ref{stretch}), for various experimental conditions. Both axes are dimensionless.
Solid grey curve shows stretched exponential fit with $\beta =0.65$,
while grey dashed curve shows single exponential with $\beta = 1$. Inset
shows temperature dependence of the best fit values of $\beta$ at
3.6 Tesla ($\bullet$) and 9.3 Tesla ($\circ$), and average 
value of the data set $\beta=0.65$ (solid line).} 
\label{recov}
\end{figure}

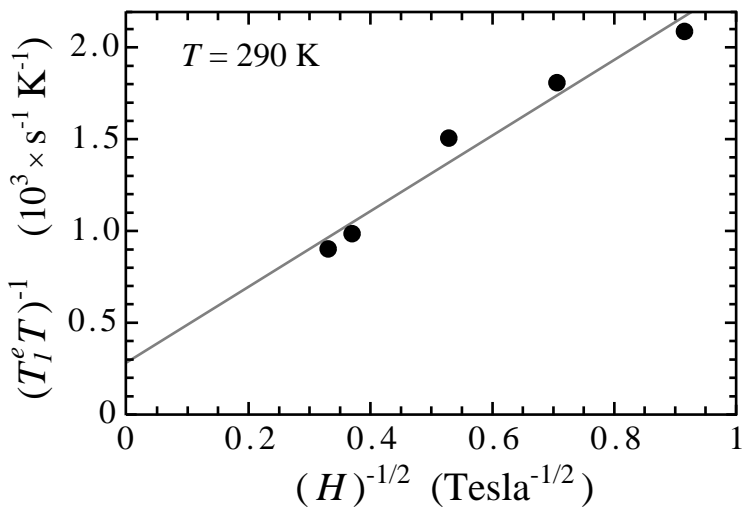
\end{document}